# Lunar Mantle Structure and Composition Inferred From Apollo 12 – Explorer 35 Electromagnetic Sounding

Robert E. Grimm[1*]

[1]Solar System Science and Exploration Division,
Southwest Research Institute, Boulder, Colorado, USA

*1050 Walnut St. #300
Boulder, CO 80302, USA
robert.grimm@swri.org

**Highlights**  *(3-5 bullets, max 85 char ea, also separate file)*

- New inversions of induction data for electrical conductivity as a function of depth
- Fit to estimated temperatures using Fe fraction (Mg#) instead of fixed-Fe minerals
- Depths 400–1200 km resolved by data <1 mHz and match thermal conduction profiles
- Mantle is well-mixed at Mg# 81 ± 10 due to incomplete overturn or prior convection
- Higher-frequency data do not yield plausible solutions: use magnetotellurics next

**Abstract**

Constraints on the interior structure of the Moon have been derived from its inductive response, principally as measured by the magnetic transfer function (TF) between the distantly orbiting Explorer 35 satellite and the Apollo 12 surface station. The most successful prior studies used a dataset spanning 0.01–1 mHz, so the lunar response could be modeled as a simple dipole. However, earlier efforts also produced transfer functions up to 40 mHz. The smaller electromagnetic skin depth at higher frequency would better resolve the uppermost mantle—where key information about primitive lunar evolution may still be preserved—but requires a multipole treatment.

I compute new profiles of electrical conductivity vs depth using both low- and high-frequency ranges of published Apollo-Explorer TFs. Using the low-frequency data, I derive temperature profiles at depths >400 km (<1 mHz) consistent with conductive heat loss and expectations of the iron (and possibly water) content of the mantle. The near-constant iron fraction (Mg# 81 ± 10) could imply efficient mixing due to now-defunct convection. Alternatively, incomplete overturn of gravitationally unstable magma-ocean cumulates could have left a heterogeneous distribution of minerals at hundred-km scales that are not resolved by electromagnetic sounding. A third explanation is that the electromagnetically probed region may be the initial equilibrium crystallization in a mantle that did not buoyantly overturn.

In contrast, the high-frequency data produced higher conductivities than expected, requiring unrealistically low Mg# or high water content at depths 200-400 km. Either the published transfer functions >> 1 mHz are incorrect, or the TF multipole method at the Moon is unreliable. Future electromagnetic sounding using the magnetotelluric method can operate up to 100s Hz and is largely insensitive to multipole effects, resolving structure to 100 km or less.

# 1. Introduction

The anatomy of the Moon was forged in the furnace of its global magma ocean. Most, if not all, of the nascent Moon was likely molten after it accreted from the debris of a giant impact to the Earth (e.g., Salmon and Canup, 2012). In addition to a small iron core, cooling of the magma ocean produced a sequence of mineral crystallization that is relatively well understood (Snyder et al., 1992; Elardo et al., 2011; Elkins-Tanton et al., 2011, Johnson et al., 2021). With iron-rich minerals forming later and higher in the cumulate pile (particularly ilmenite-bearing cumulates or IBCs), the earliest solid interior was gravitationally unstable, which drove overturn (Hess and Parmentier, 1995). The extent of homogenization of the interior imposed by overturn is not established, which could vary from a stable layered structure (Elkins-Tanton et al., 2011), to heterogeneous residuals of Rayleigh-Taylor instabilities (Zhao et al., 2019), to a fully mixed mantle following thermally driven convection (Zhang et al., 2013). The profile of the atomic ratio of magnesium to iron plus magnesium (Mg#) is one discriminant of these configurations. Geophysical imaging of the interior can constrain structure, composition, and temperature, and thus continue to advance our understanding of the Moon as an archetype for the early evolution of terrestrial planets.

Four magnetometers were placed on the surface of the Moon during the Apollo missions (see Dyal et al., 1974, for a review). Although static fields were analyzed for all, the time-varying records from the Apollo 12 Lunar Surface Magnetometer have special significance because similar data from the Explorer 35 satellite were obtained simultaneously. Synchronized comparison of magnetic fields at the surface and distant from the Moon has been the foundation of lunar electromagnetic (EM) sounding (see Sonett, 1982, for a review).

Time-varying magnetic fields induce electrical currents in planetary interiors according to Faraday's Law. The secondary magnetic fields associated with these eddy currents can be detected at the surface. The ratio of the surface field to the distant field is the magnetic transfer function (TF) and measures the sum of source and induced fields normalized by the source field. The time- or frequency-dependent TF is determined by the electrical-conductivity profile of the interior, which in turn depends on temperature and composition. The goal of EM sounding, then, is to derive interior properties from induction responses such as the transfer function.

Hood et al. (1982) produced Explorer 35-to-Apollo 12 frequency-domain TFs spanning 0.01 to 1 mHz (tabulated in Hobbs et al., 1983). Lower frequencies penetrate deeper according to the skin-depth effect and so inversion techniques can reconstruct the profile of electrical conductivity vs. depth. These data were further modeled and interpreted by Hood and Sonett (1982) and Khan et al. (2006, 2014). However, the upper frequency limit implies an effective minimum depth ~400 km (see below). Higher frequencies are necessary to resolve the upper mantle <400 km, which may still preserve structure related to early lunar vertical and lateral differentiation (e.g., Elkins-Tanton et al., 2011; Moriarty et al., 2021; Wieczorek and Phillips, 2000; Grimm 2013).

Sonett et al. (1972) produced frequency-domain TFs over the interval 0.5 to 40 mHz. In principle, depths as shallow as 200 km (see below) or less could be imaged. While the shorter time series were more tractable than the long swaths treated by Hood et al. (1982), the resulting TFs have considerable scatter. Modeling is also more challenging because multipole responses must be included, which in turn require additional estimated parameters. The rudimentary conductivity structures (some incorporating unphysical high-conductivity layers) presented by Sonett and colleagues have limited utility today, although Hobbs (1977) derived a smooth profile over the interval ~100-600 km depth.

The first objective of this paper is to merge the "low-frequency" (LF) transfer functions from Hood et al. (1982) and the "high-frequency" (HF) transfer functions from Sonett et al. (1972) (note that the use of LF and HF here are formally below the International Telecommunication Union extremely low frequency ELF band). Separate LF and HF inversions are performed for comparison with the "all-frequency" (AF) merged data. The second objective is to model these derived conductivities in terms of several estimated temperature structures in the literature and laboratory measurements of mineral conductivity. Finally, the results are assessed for vertical compositional variations within the Moon.

## 2. Background

*2.1. Magnetic Transfer Functions*

The magnetic transfer function is:

$$A = \frac{B_{surf}}{B_{dist}} = \frac{B_s + B_i}{B_s} \qquad (1)$$

where $B_{surf}$ is the magnetic field measured at the surface of the Moon and $B_{dist}$ is that measured at a distant location. This also defines the relationship between the source field $B_s$ and the induced field $B_i$. A component-by-component analysis is implicit in Eqn. (1). The spherical coordinate system $(r,\theta,\phi)$ is defined with $\theta = 0$ at the anti-incidence point of the incoming wave vector (Sonett et al., 1972, their Fig. 6). Because both the wave vectors and the lunar equator lie near the ecliptic plane, the east-west and north-south components of the TF are good approximations to $A_\theta$ and $A_\phi$, respectively.

Transfer functions can be evaluated in the time or frequency domain and in any of the space environments around the Moon (solar wind, magnetosheath, magnetotail, or wake, with both measurements in the same environment). Although these parameters are theoretically independent, in practice workers chose frequency domain for the wind/sheath (e.g., Sonett et al., 1972; Hood et al., 1982) and time domain for the wake (e.g., Dyal et al, 1976). Overall, the two approaches yielded comparable results, with the wake results offset to higher conductivity (see Fig. 1 of Hood and Sonett, 1982). This may be due to confinement of induced fields in the wake (Fatemi et al., 2015). An unconfined vacuum response was assumed in modeling induction responses there; if the transfer function is biased high, the derived conductivity will follow. Transfer functions in the frequency domain were constructed from many time series and the resulting averaged tabulations can be readily used by others, whereas only a handful of individual transients were used in the time-domain analyses. While others have begun pursuing better understanding of wake induction responses in the time domain (Fuqua Haviland et al., 2018), I use only frequency-domain data obtained in the solar wind and magnetosheath (**Fig. 1**).

The data-reduction methods used by Sonett et al. (1972) and Hood et al. (1982) were broadly similar, namely Fourier transforms of a number of time series ("swaths") and formation of transfer functions from the ratios of the square roots of the power spectral densities (PSDs) of the two measurements. This approach loses the phase, an important complementary measurement to amplitude, and may introduce bias. Timing accuracy was insufficient then to determine complex responses, but most contemporary electromagnetic frequency-domain methods treat both phase and amplitude.

Sonett et al. (1972) compiled 31 swaths with varying degrees of overlap over the first three lunations at the Apollo 12 site. These data also had to satisfy requirements that Explorer 35 was near apoapsis and in the same space environment. The maximum frequency was 40 mHz, conservatively half of the Explorer 35 Nyquist frequency. The minimum frequency 0.5 mHz followed from a requirement of at least 15 cycles of the longest record available.

Plasma of the solar wind and magnetosheath is sufficiently energetic to confine induced magnetic fields within the lunar surface (Blank and Sill, 1969). This forces the radial component of the transfer function $A_r$ to unity at all frequencies and amplifies the tangential (horizontal) components compared to a free-space response. In theory (Schubert and Schwartz, 1972), the two tangential components $A_\theta$ and $A_\phi$ differ only at higher frequencies due to multipole contributions. However, $A_\phi \approx A_\theta + 0.4$ across all frequencies in Sonett et al. (1972; their Fig. 1). They ascribed this to plasma noise and bias, and attempted to rectify it by forming a minimum $A_{min}$ by principal-component rotation of the tangential components at each frequency (their Fig. 3).

Hood et al. (1982) focused on extending the transfer-function analysis to lower frequencies. They carefully winnowed Apollo-Explorer time series in several steps, finally selecting seven swaths from the first two Apollo 12 surface lunations with durations 32–52 hours. Hood and colleagues computed the TF as the square root of the slope of the linear best fit to the Explorer vs Apollo PSDs. They selected $A_\phi$ for further analysis as having higher signal-to-noise. Even so, the lowest reported frequencies 0.01–0.02 mHz have error bars that transgress the minimum allowable transverse transfer function $A_\phi, A_\theta \geq 1$. This may be due to intrinsic noise, insufficiently long time series, or movement of Explorer 35 into the magnetotail. The upper frequency limit in the analysis of Hood et al. (1982) was 1 mHz.

*2.2 Forward Modeling of Conductivity Structure*

In order to solve for conductivity structure, a forward model must compute TFs for any input conductivity profile. The frequency domain tangential magnetic transfer functions $A_\theta$ and $A_\phi$ for a Moon with an arbitrary radial conductivity profile and fully confined by the solar wind were derived by Schubert & Schwartz (1972). These multipole equations were reproduced by Sonett et al. (1972) and Hobbs (1977) and are not repeated here. In summary, each degree $\ell$ is the product of geometrical terms for spherical coordinates and a function that depends on the solution of the wave-diffusion (Helmholtz) equation within the Moon and evaluated at the surface. The radial dependence of electrical conductivity appears in the diffusion term (wave behavior can be neglected below 100s Hz). Sonett et al. (1972) summed the responses through degree five.

Because the incident fields are orthogonal to the propagation/contact direction, only the (asymmetric) order-one terms are included. In addition to the selected conductivity profile, the solution depends on $\theta$ of the surface station and the solar-wind velocity $v_p$. The latter is used to determine the wavelength at each frequency $\lambda = v_p/f$, where $f$ is the frequency. These are free parameters: Sonett et al. (1972) found best fits to the HF data at $\theta = 150°$ and $v_p = 200$ km s$^{-1}$. These parameters were also adopted by Hobbs (1977).

*2.3 Electrical Conductivity of Minerals*

The electrical conductivity of minerals deep in planetary interiors is determined by point-defect chemistry, that is, imperfections in the crystal lattice (see Tyburczy, 2007, Yoshino, 2010, and Karato and Wang, 2013, for reviews). There are three principal mechanisms, which depend strongly on temperature. Ionic conduction is caused by the mobility of charged cation vacancies. In mafic silicates, ionic conduction is mainly due to magnesium and tends to dominate at high temperatures, because of higher activation energy. Substitution of $Fe^{3+}$ on an $Mg^{2+}$ site generates a mobile electron and electron hole; their complementary movement is known as hopping or small-polaron conduction. Finally, protons from dissociated water are highly mobile in silicates and such conduction tends to dominate at low temperatures, due to lower activation energy. Yoshino (2010; their Fig. 12) schematically illustrates the interaction of these three mechanisms on an Arrhenius plot. The three mechanisms operate in parallel and so conductivities can be added linearly.

Lunar mantle mineralogy has been calculated from estimated bulk composition based on equilibrium thermodynamics (BVSP, 1981; Khan, 2006; 2014; Kuskov et al., 2019, and references therein) and the magma-ocean crystallization sequence (Snyder et al., 1992; Elkins-Tanton et al., 2011; Johnson et al., 2021). Olivine and orthopyroxene likely comprise >80% of the lunar mantle, with clinopyroxene, garnet, and ilmenite making up most of the remainder. An established approach to modeling electrical conductivity is to select one constitutive relation per mineral and then—usually with additional constraints—derive the proportion of each mineral (e.g., Khan et al., 2006; 2014). These conductivities were measured at fixed iron contents (i.e, on a single specimen each). Appendix A presents a new synthesis of conductivity formulae emphasizing the effect of variable iron content within individual minerals.

Ilmenite-bearing cumulates (IBCs) are expected to have crystallized late in the magma-ocean sequence and sunk to the base of the mantle. Yet models by Zhao et al. (2019) indicate that cooling and solidification could have trapped a substantial fraction of IBC in the uppermost mantle, with some residual distributed throughout the mantle. FeTiO$_3$ is much more conductive than ferromagnesian silicates and could strongly influence electrical conductivity (see Appendix A). However, the total abundance of ilmenite and other conductive oxides is only ~1% of the magma-ocean volume, expressed as ~10% volume of the last-crystallizing ~10% (e.g., Snyder et al., 1992; Elkins-Tanton et al., 2011). At such low concentrations, the ilmenite grains are unlikely to be interconnected. Hence, the conductivity would fall near the lower Hashin-Shtrikman bound and can be neglected. Interconnected ilmenite would increase the conductivity of the uppermost mantle by orders of magnitude and would have a profound effect at all frequencies that is inconsistent with the observed transfer functions.

The Moon may possess a zone of partially molten silicates surrounding an iron core (Weber et al., 2011). Partial melting of the mantle, particularly IBCs, can greatly increase electrical conductivity (e.g., Khan et al., 2014), and the conductivity of an iron core would be much larger than any silicates. Rigorous analysis of the low-frequency TF of Hood et al. (1982) allows for a conductive core <435 km radius (>1300 km depth; Hobbs et al., 1983). Asymptotic induction studies using Lunar Prospector and Kaguya (Hood et al., 1999; Shimizu et al., 2013) yield a similar maximum radius. However, these studies did not have the frequency response to distinguish metal from partially molten silicates (Grimm and Delory, 2012). The work presented here does not extend to the core, whether iron or silicate.

*2.4 Lunar Interior Temperatures*

Because electrical conductivity depends on both mineralogy and temperature, independent constraints must be used to separate the two factors. Heat-flow measurements are highly complementary to EM sounding for temperature, and future geophysical exploration of the Moon optimally pairs these methods (e.g., Neal et al., 2021, Nagihara et al., 2022; Grimm et al., 2023). Here, I use published estimates of the present interior temperature profile of the Moon (**Fig. 2**) and solve for compositions that match electrical conductivity.

Gagnepain-Beyneix et al. (2006) derived a comparatively cold interior by comparing seismic constraints to an internally heated thermal-conduction model. Grimm (2013) calculated somewhat hotter temperatures with a similar thermal model, which was taken as the "background" distant from anomalous heating assigned to the Procellarum KREEP Terrane (PKT; cf. Wieczorek and Phillips, 2000). In a remarkable early 2D convection model, Toksöz et al. (1978) found present-day temperatures intermediate between these models, as convection has essentially stopped. A 1D parameterized-convection model by Schubert et al. (1979) remains in vigorous convection today, as evidenced by the classic linear slope in the stagnant lid over a nearly isothermal (adiabatic), convecting interior. Recent 3D convection modeling by Zhang et al. (2013) yields comparable profiles, at somewhat colder or hotter interior temperatures depending on the activation energy of viscosity. The agreement of the Schubert model is partly due to a fortuitous (shrewd?) choice of a 400-km-thick lid, whereas this feature evolves naturally from the Zhang model. The hottest model by Zhang and colleagues approaches, but does not cross, the solidus (Ringwood, 1976).

Using both electrical-conductivity and geodetic-tidal constraints, Khan et al. (2014) obtained hotter temperatures than any of the conduction models, with a nearly linear profile from 20-1200 km depth. Temperatures are likely high because of the erroneous dominance of clinopyroxene in the electrical conductivity (due to high iron content and high $fO_2$ of the reference measurements; see Appendix A). A linear profile is challenging to reconcile with either convection or conduction. Furthermore, the slope 0.6 K/km implies heat flow of only 2 mW/m$^2$ at an average 3.3 W-m/K, well below global estimates (e.g., Warren and Rasmussen, 1987; Siegler and Smrekar, 2013). The temperature profile of Khan et al. (2006) is more complex but averages nearly 500 K lower than their 2014 result, and ~200 K lower than that of Gagnepain-Beyneix and colleagues.

## 3. Methods
### *3.1. Merged Data*

I sought to merge the Hood (LF) and Sonett (HF) datasets into a single tangential TF spectrum spanning 0.01 to 40 mHz and so derive lunar interior properties over the widest possible range of depths. First, the arithmetically spaced LF data as tabulated in Hobbs et al. (1983) were interpolated onto a log-spaced grid at 9 points/decade. These data are formally $A_\phi$ but in the dipole limit are equivalent to $A_\theta$. Next, the $A_\phi$ HF data were digitized, smoothed, and also interpolated to log spacing at 9 points/decade. Standard errors were specified for each frequency in the LF data (Hobbs et al., 1983); representative errors shown for some HF points (Sonett et al., 1972) were scaled over all HF data. A correction of +0.1 units was applied to the HF data in order to match the LF data in the overlap region 0.5-1 mHz. For simplicity, I refer to $A_\phi$ from here on simply as the tangential tranfer function $A_T$.

The merged transfer function (**Fig. 1**) increases with frequency due to solar-wind compression and increased screening by eddy currents. When the wavelength of the plasma turbulence is comparable to the lunar radius (at frequencies of several mHz), the response can no longer be described as a simple dipole; multipoles cause the HF rollover of the TF. The slope shallows considerably as the low-frequency limit $A_T \to 1$ is approached near 0.01 mHz; this indicates that EM fields are nearly fully penetrating the Moon.

*3.2. Penetration Depth*

The electromagnetic inductive length scale or penetration depth $C$ is classically defined for the Earth in terms of the divergence equation or the wave impedance (Schmucker, 1970), but the dipole response is simply $C = a/2A_T$, where $a$ is the lunar radius (Weidelt, 1972; Hobbs et al., 1983). Higher $A_T$ at higher frequency is associated with smaller $C$, in accordance with the skin-depth effect, and indeed $C$ is $1/\sqrt{2}$ times the EM skin depth, i.e., $C = 1/\sqrt{\omega\mu\sigma}$, where $\omega$ is the angular frequency, $\mu$ is the magnetic permeability, and $\sigma$ is the (apparent) conductivity.

The low-frequency asymptote $A_T \to 1$ implies $C \to a/2$. At face value, this suggests that the maximum EM penetration depth is just half of the lunar radius. However, $a/2$ simply represents the depth of an equivalent conductor that gives the correct spherical response. For direct interpretation, $C$ must be mapped to cartesian coordinates using Weidelt's (1972) algebraic transformations. These equations are reproduced in Hobbs et al. (1983) and Grimm and Delory (2012). In fact, both papers used the Weidelt transformations to solve the spherical-induction problem using equivalent cartesian coordinates. The transformed inductive length scale $C'$ varies over depth 0 to $a$.

$C$ is commonly used in asymptotic inversions, which provide a rough approximation to the conductivity profile using a one-to-one mapping of frequency to depth. In practice, asymptotic inversions of synthetic conductivity profiles indicate that an empirical correction factor 0.8-0.9 brings the recovered depths into alignment with depths recovered from full inversions (see Whittal and Oldenberg, 1992, for a review). I found a correction factor 0.85 best matches asymptotic inversions for synthetic lunar profiles, so the effective penetration depth is taken as $D$

= 0.85$C'$. I use $D$ extensively in describing EM length scales as a function of frequency, but it must always be remembered that this is an asymptotic approximation.

**Figure 3** shows the original $C$ and the Weidelt-transformed and empirically adjusted $D$. The LF data correspond to penetration depths ~400–1200 km. Hobbs et al. (1983) used an alternative asymptotic approach to derive a penetration depth of 1300 km. The lowest-frequency signals effectively sense no deeper than the upper limit, and the highest-frequency signals cannot resolve the zone above the lower limit. Conversely, $D$ = 200–550 km for the HF data. Of course, inverse methods will yield conductivity profiles beyond these bounds, but they are increasingly driven by the boundary conditions. Therefore, these approximate bounds will be obeyed when plotting and interpreting the sounding results derived here.

*3.3. Inversion for Conductivity Structure*

The solution for a conductivity profile from the TF was performed using the classical Levenberg-Marquardt method as implemented in Matlab® `lsqnonlin`, which minimizes the sum of the squares of the misfit between the observed and predicted data. Inversions were carried out separately for the LF, HF, and all-frequency (AF) data. Following limited sensitivity testing with the HF data, I selected $\theta$ = 150° and $v_p$ = 200 km/s, following Sonett et al. (1972) and Hobbs (1977). The calculations were truncated at $\ell$ =3, as no significant changes were observed at higher degree.

The model vector comprised 29 layers with thickness 50 km from 0 to 1400 km and was specified as the logarithmic increment to conductivity of the previous layer. By allowing only positive increments going downward from a specified surface value $\sigma_{top}$, the conductivity is forced to increase monotonically with depth. Conversely, the conductivity at the bottom of the model $\sigma_{bot}$ can be fixed, and only negative increments allowed going up. A conductivity profile that monotonically increases with depth is plausible for the deep interior of the Moon and is naturally smooth, obviating the need for any roughness-based regularization (e.g., Constable et al., 1987).

After iteratively adjusting the boundary conductivities, only small differences remained between the top-down and bottom-up approaches. I selected top-down for HF with $\sigma_{top}$ = 3x10$^{-5}$ S/m and the bottom-up for LF and AF with $\sigma_{bot}$ = 0.06 S/m. The latter approach eliminates large lower-boundary conductivity excursions that may or may not be physical. Note that these should values should not be interpreted literally as they lie outside the valid depth bounds of the data and function solely to regulate the behavior of the in-bounds solution.

The best-fitting model and its errors were determined from ensemble statistics. Forty-five trial TFs were generated by sampling the observed TF assuming that the tabulated errors are one standard deviation of a normal distribution. The 45 resulting conductivity profiles were ordered at each depth and the error envelope reported as the 68$^{th}$ percentile. Forty-five trials exceed the conventional 30 to satisfy the central limit theorem (so the 68$^{th}$ percentiles should approximate the standard deviation of a normal distribution) and conveniently reject ±7 profiles as outliers.

*3.6. Temperature and Composition*

I selected the temperature profiles of Gagnepain-Beyneix et al. (2006) and Grimm (2013) as representative of "cold" vs "hot" conduction models. I also treated the convective temperature profiles of Zhang et al. (2013), but the constant temperature in the convecting zone is inconsistent with the steady increase of conductivity with depth.

The recent trend in electrical-conductivity experiments is to improve the constitutive relations to account for variable iron or water content by analyzing multiple specimens (e.g. Yoshino, 2008; 2009; Wang et al., 2014; Verhoeven and Vacher, 2016). However, it is beyond the scope of this study to attempt to derive both mineral proportions and their variable compositions. Instead, I model the lunar mantle using one mineral at a time but allowing vertically variable iron or water content. Iron content is expressed relative to molar iron + magnesium either as a fraction $X_{Fe}$ or as the complementary magnesium number, Mg# = 100(1–$X_{Fe}$). Water content $c_w$ is in traditionally in percent but can also be given as parts-per million by weight (ppmw = $10^4 c_w$). Details of mineral conductivities are given in Appendix A. Unique solutions follow for $X_{Fe}$ or $c_w$ as a function of depth when a temperature profile, such as those described above, is specified.

Due to uncertainty in the electrical conductivity of clinopyroxene (Appendix A), only olivine and orthopyroxene were ultimately fit to the conductivity structure. These two minerals dominate the composition of the lunar interior.

**4. Results**

*4.1 Conductivity vs Depth*

**Fig. 4** shows model solutions and data fits for each of the LF, AF, and HF bandwidths. The LF result is in excellent agreement with Khan et al. (2014). Both profiles lie within the conductivity limits of Hood et al., 1982, albeit near the upper limit. Recall, however, that asymptotic inversion (Fig. 3) indicates that the LF data—the same as used by Khan, Hood, and others—produces well-constrained results only over the depth range ~400–1200 km. The LF profile over this interval can be fit well to the function $\sigma[S/m] = 1.76 \times 10^{-4} \exp(z[km]/210)$ ($r^2 = 0.994$).

The recovered HF profile has a shallower semilog slope 200-400 km and a steeper slope 400-550 km compared to the LF slope. The shallow segment is in excellent agreement with Hobbs (1977), but rapidly diverges below. In the overlap region, the conductivity of the HF profile trends sharply across LF profile.

The derived AF conductivity profile shows three distinct regions. The AF profile matches the LF semilog trend 900-1200 km and extends it 200-400 km. However, at 400-900 km, the AF profile is up to a factor of 2 lower conductivity than the LF result.

*4.2 Composition from Electrical Conductivity and Temperature*

Iron fractions $X_{Fe}$ in olivine (ol) and orthopyroxene (opx) for the thermal-conduction models of Gagnepain-Beyneix et al. (2006: GB06) and Grimm (2013: Gr13) are shown in **Fig. 5** for the median LF electrical-conductivity solution and in **Fig. 6** for the median HF solution. **Table 1** gives volume-weighted means and standard deviations of the iron fractions as the complementary Mg# for easier comparison to the literature. The final entries propagate the errors from the corresponding electrical-conductivity solutions to the standard deviation calculated over all four models (the variance among the thermal-composition models is much larger than the variance in the electrical-conductivity solutions).

The LF profiles are nearly constant with depth, with standard deviations just several percent of the mean. Differences between the thermal models cause the largest differences in the curves: higher $X_{Fe}$ (lower Mg#) is derived for GB06 compared to Gr13 because higher mineral conductivity is required to offset the lower temperature. Orthopyroxene is generally more conductive than olivine, except at very low $X_{Fe}$. Averaging all four models together yields Mg# $81 \pm 10$.

In contrast, iron profiles for the HF conductivity solution can vary by more than a factor of three, with $X_{Fe}$ increasing with decreasing depth to values >0.6 (Mg# <40). The Mg# 63 ± 17 averaged over all four models is unreasonably low for the Moon, even allowing for preservation of a primordial iron-rich upper mantle (see Discussion). Alternatively, $H_2O$ can provide the needed electrical conductivity This also results in a profile in which the conductive constituent increases strongly with decreasing depth. **Table 2** gives volume weighted average $H_2O$ content and standard deviation, fit to the HF solution at constant Mg# 81. The large mean water abundances >1000 ppmw and the vertical distribution are also implausible. The HF-derived conductivities are therefore likely too high (see Discussion) and so the HF solution yields no useful insights into the lunar interior. For this reason, compositional profiles for the AF conductivity solution are not shown, although summary statistics are given in Tables 2 and 3.

## 5. Discussion

*5.1 Inverted Conductivity Structure*

As noted above, the new LF inversion for conductivity structure 400-1200 km depth closely follows that of Khan et al. (2014) and lies within the error bounds of Hood et al. (1982), whereas the HF and AF inversions introduce different, apparently incompatible, structure. In this section, I consider factors influencing the EM inversions and the length scales that the 1D inversions sample. The latter is also relevant to a recent attempt at global electromagnetic sounding by Mittelholz et al. (2021).

As described above, the conductivity at the top of the model was empirically determined to be ~$3 \times 10^{-5}$ S/m. However, the true near-surface conductivity is likely far lower. Dyal and Parkin (1977) used limits to the toroidal induction mode to infer an upper limit to near surface conductivity <$10^{-8}$ S/m at depth <80 km. Bulk conductivities of lunar rocks at temperatures <300 K are very small, <$10^{-10}$ S/m (Carrier et al., 1991). Therefore, there must be a much stronger gradient between the top of the LF model (~$10^{-3}$ S/m at 400 km) and the true surface conductivity than is represented in the HF result. The inversion algorithm can recover synthetic profiles with a low-conductivity segment. Therefore, possible explanations for this discrepancy include data quality and model assumptions.

Data quality is an obvious potential shortcoming. Sonett et al. (1972) had a limited number of swaths available and devoted a large portion of their paper to attempting to correct for plasma artifacts. Hobbs (1977) is the only other published attempt to invert these data. Success of a future TF experiment would therefore depend on better data integration or processing from longer observations and on the ability to remove plasma signatures.

The spherically symmetric model assumes global confinement by the solar wind, as well as knowledge of the solar-wind velocity and colatitude from the wave-impact point. None of these assumptions have been rigorously tested in this context. Furthermore, departure from spherical symmetry is a possibility, which would distort the one-dimensional interpretation. The Apollo 12 site is ~800 km from the nominal PKT boundary (Jolliff et al., 2000), possibly separating terranes of hugely different interior properties (e.g., Wieczorek and Phillips, 2000).

EM sounding is sensitive laterally to a distance $L$ of about one skin depth (Vozoff, 1991), or ~1.7$D$ given in Fig. 3. This was established by assessing 2D cartesian models with a lateral discontinuity. A boundary at 800-km distance should influence a 1D interpretation at $D > 470$ km or frequencies <0.6 mHz in the asymptotic mapping. This lies within the LF-HF overlap and may be a plausible indication of the initial contact with a boundary. However, the transition between two different regions of uniform conductivity (in layered or lateral contact) requires 2-3 decades of frequency (e.g., Vozoff, 1991). Then the entire LF profile would be biased, which is inconsistent with results presented here and in the literature for lunar interior structure. Although it is conceivable that detailed 2D modeling could reveal some special case compressing the transition bandwidth and producing the required conductivity structure, I tentatively conclude that conductivity structure incorporating the HF data is not consistent with the influence of lateral heterogeneity.

The asymptotic mapping of frequency to distance can also be used to assess the volume of the lunar interior sampled by EM sounding. Using the same relationship $L \sim 1.7D$, the full 200-1200 depth interval is laterally sensitive to $L = \pm 300$-2000 km. Using solid angles, the global coverage fraction is approximately $\alpha = \pi L^2 / 4\pi(a-D)^2$, which breaks down (say $\alpha > 0.5$) for $D > 800$ km. In other words, any EM sounding attaining depths >800 km can be treated as approaching a global average at those depths. On the other hand, depths of 200 and 400 km sample just 1% and 6%, respectively, of the sphere.

Seeking a global EM sounding result, Mittelholz et al. (2021) presented an innovative application of classical Geomagnetic Depth Sounding (GDS) to the Moon using low-altitude orbital magnetic data. GDS (Banks, 1969; see Constable, 2007, for a review) recovers the inductive length scale $C$ from the ratio of the vertical to horizontal magnetic field as a function of spherical-harmonic degree and order. Mittelholz and colleagues derived $C$ from the dipole response of the Moon in the quasi-uniform field in the magnetotail over the frequency range $6 \times 10^{-6}$ to $4 \times 10^{-5}$ Hz. However, $C$-values for frequencies $<1.3 \times 10^{-5}$ Hz either exceed the $a/2$ limit (see above) or have large error bounds that substantially breach the limit, and therefore are unphysical. Higher frequencies are in reasonable agreement with Apollo-Explorer LF data with corresponding asymptotic depth sensitivity $D = 900-1200$ km. Per the coverage analysis above, EM sounding by any method is attaining global scale here in the deep mantle, but mid-to-upper mantle structure is still local as it is determined by the Apollo-Explorer data. Mittelholz et al. (2021) have provided an important proof-of-concept, but GDS for global properties of the Moon remains to be demonstrated.

The origin of the differences between the LF and HF/AF inversions remains unresolved. Agreement of the LF inversion with prior work suggests an issue with the HF data. Finally, modeling the HF data above several mHz requires a multipole formulation and associated parameter assumptions. This is likely to be the fundamental limit to the transfer function method. For comparison, terrestrial GDS is routinely applied using the ring-current dipole only (e.g., Banks, 1969; Olsen, 1999; Kuvshinov et al., 2012); attempts to use multipoles of the solar-quiet ionosphere variation (Bahr and Filloux, 1989) have met with limited success.

*5.2 Vertical Temperature and Composition*

Due to the uncertainties in the HF data and inversion, I restrict discussion to the LF results. The mean Mg#s derived here for the lunar mantle 400-1200 km depth vary from 79 to 86 assuming an olivine composition and from 70 to 87 assuming an orthopyroxene composition, as applied to two the different thermal-conduction models. Over all four scenarios, the volume-weighted Mg# is $81 \pm 10$, including the inversion error bounds. This is in good agreement with the green-glass source models for the lunar mantle of Longhi (2006), Mg# 80-81. Mg#s vary from 80 to 84 ($82 \pm 2$) over eleven magma-ocean bulk compositions tabulated by Elkins-Tanton et al. (2011).

The electrical conductivity profile derived here (and supported by previous studies, Hood and Sonett, 1982; Khan et al., 2006; 2014) is consistent with a nearly uniform mantle >400 km depth. Variations of Mg# within any one of the four essential models (ol vs opx for two different thermal-conduction models, Table 1) are only 4–13 points. The methodology of Khan et al. (2006; 2014) yields essentially no variation in Mg#: the linear geotherm results in a temperature profile that can exactly track the Arrhenius conductivity relations, so the preferred mineral assemblage (each with constant Mg#) is nearly constant with depth. Also note that, in a comprehensive review, Garcia et al. (2019) concluded that there is no clear evidence for a mid-mantle (600-1200 km) seismic discontinuity. For seismology, a uniform mantle remains the null hypothesis.

Initial equilibrium crystallization of the magma ocean produces ~50% vol olivine or orthopyroxene. This is followed by fractional crystallization that leads to the ferroan anorthosite crust and an iron-enriched upper mantle, including IBCs. The upper limit to the equilibrium/fractional boundary is ~360 km, for a whole-Moon magma ocean (Johnson et al., 2021). Buoyancy-driven overturn is expected but may have been retarded (Zhao et al., 2013; Moriarty et al., 2021; Zhang et al., 2023). If the primordial stratigraphy was preserved, the LF data would only be sensitive to the early, nearly uniform olivine or orthopyroxene zone, as observed (Fig. 5). The HF-inferred composition does trend toward higher iron fraction at shallower depth, qualitatively tracking fractional crystallization, but the recovered $X_{Fe}$ is much too large (Fig. 6). The HF data may be rejected due to likely bias (see above), but the possibility of a preserved original magma-ocean stratigraphy cannot be ruled out. On the other hand, the gravitationally stable, overturned cumulate pile can be ruled out because the decreasing volume going deep into the interior results in strong gradients in iron content (Elkins-Tanton et al., 2011) that are not observed.

Alternatively, the lunar mantle may have been previously well-mixed, at least at the Apollo 12 site at lateral scales exceeding ±700 km implied by the minimum 400-km depth in the LF inversion. This scale of geophysical homogeneity still permits smaller-scale geochemical heterogeneity, e.g., the source regions of high-Ti basalts. Incomplete mantle turnover (Zhao et al., 2019) could also result in mixing that would appear nearly homogeneous at the EM-averaging scale of hundreds of kilometers. Present-day vigorous convection (e.g., Zhang et al., 2013) is ruled out, although prior convection that has become sluggish may mimic conduction (Toksöz et al., 1978).

If a uniform mantle Mg# = 81 is enforced, residual conductivity variations can be modeled as $H_2O$. Table 2 was previously used to demonstrate the unreasonably large water contents required by the AF/HF conductivity models, but the LF requirements are very modest, from a few to a few hundred ppmw depending on the scenario and constitutive relation. Karato (2013) derived 10-100 ppmw for the lunar mantle. Appendix A recounts how Karato's formula requires significantly less water for a specified conductivity than does the formula of Yoshino et al. (2008). Furthermore, Karato's comparison of predicted conductivity to geophysical observations is highly generalized and does not produce the relatively close fits presented here. Nonetheless, O(100 ppmw) $H_2O$ is consistent between this work and theirs and has ample support in modern lunar geochemistry (Saal et al., 2008; Robinson and Taylor, 2014; Hauri et al., 2015).

## 6. Conclusion

Magnetic transfer functions between the remote Explorer 35 satellite and the Apollo 12 surface station include the inductive response of the Moon, which can be inverted for a 1D conductivity profile. I reanalyzed the transfer functions 0.01–1 mHz from Hood et al. (1982) and derived a conductivity profile within the error bounds of that early work, but in close agreement with the profile of Khan et al. (2014). The conductivity is well-fit by a simple exponential depth dependence and is sensitive to depths approximately 400-1200 km. Constitutive relations for electrical conductivity as a function of mineralogy and temperature can be fit to the depth profile. I chose to match previously published temperature profiles with a single mineral at a time, but solving for the best-fitting iron or water fraction. Bounding the results using olivine and orthopyroxene and the conductive temperature profiles of Gagnepain-Beyneix et al. (2006) and Grimm (2013), I derived Mg# (the complement of iron fraction) of 81 ± 10 over this depth interval. With Mg# fixed at 81, residual conductivity can be fit with an average of 10s ppmw $H_2O$. The near-constant Mg# 400-1200 km is most simply explained by prior mantle mixing, but could also represent initial equilibrium crystallization in a mantle that did not overturn.

Sonett et al. (1972) produced magnetic transfer functions 0.5–40 mHz. I merged these "high-frequency" transfer functions with the prior "low-frequency" ones in order to determine if shallower structure and composition could be resolved. Because turbulence wavelengths in the solar wind become comparable to the Moon's radius with increasing frequency, a multipole treatment is required. This also requires introduction of additional parameters. Electrical conductivity recovered from the high-frequency data at 200-550 km depth was much higher than expected, which in turn requires unreasonably high iron or water content. These data did not usefully constrain the interior properties of the Moon.

I conclude that magnetic-transfer function method may be fraught at frequencies >> 1 mHz, which will limit the ability of future electromagnetic sounding to resolve the critical zone of the uppermost mantle. Alternatively, the magnetotelluric method (MT) can be used, which derives conductivity structure by jointly measuring magnetic and electric fields (e.g., Vozoff, 1991; Simpson and Bahr, 2005; Chave and Jones, 2012). As it does not require a distant reference, MT is largely insensitive to multipole effects (Grimm and Delory, 2012). A miniature MT instrument has been selected for two lunar missions (Grimm et al., 2023) and seeks to perform soundings to 1 Hz or higher, thus improving resolution of the uppermost mantle to 100 km or less.

## 7. Data Availability

The three datasets in Fig. 4 are available at [permanent URL].

## 8. Acknowledgements

This work was supported by NASA grants NASA 80MSFC20C0024 (LSITP) and 80NSSC22M0065 (PRISM). I thank Lon Hood helpful discussions, and Amir Khan and an anonymous reviewer for comments that improved the paper.

## 9. Appendix: Mineral Electrical Conductivities

The electrical conductivity of a ferromagnesian mineral can be written

$$\sigma = \sigma_0^i \exp\left(-\frac{H_i}{kT}\right) + \sigma_0^h \left(\frac{X_{Fe}}{X_{Fe,0}}\right)^{a_h} \exp\left(-\frac{H_h - b_h X_{Fe}^{1/3}}{kT}\right) + \sigma_0^p \left(\frac{c_w}{c_{w,0}}\right)^{a_p} \exp\left(-\frac{H_p - b_p c_w^{1/3}}{kT}\right)$$

(A1)

where the subscripts $i$, $h$, and $p$ refer to the ionic, $Fe^{3+}$ hopping, and proton conduction, respectively (e.g., Yoshino et al.., 2009; Karato, 2013; Verhoeven and Vacher, 2016). The standard Arrhenius parameters for reference conductivity and activation energy are $\sigma_0$ and $H$, respectively. The pre-exponential factors for impurity concentration are $X_{Fe}$, the iron fraction expressed relative to total atomic iron and magnesium, and $c_w$, the water concentration in percent. The iron fraction is normalized by a reference value $X_{Fe,0} = 0.1$ (Verhoeven and Vacher, 2016) and is taken to the $a_h$ power. The water percentage is normalized by $c_{w,0} = 0.01$ wt% in Karato (2013), but no normalization factor is used by Yoshino et al. (2009) (the offset for $c_{w,0} = 1$ moves to the $\sigma_0^p$ term). The corresponding exponent for water concentration is $a_p$. Activation energies for both impurities are decreased by factors $b$ multiplying the cube root of concentration: this is essentially a statement that activation energy decreases with smaller defect spacing. Pressure dependence is negligible in the Moon, so activation volume is discarded.

I adopt the parameters for conductivity as a function of temperature and iron content of olivine, orthopyroxene, clinopyroxene, and garnet synthesized by Wang et al. (2014) and Verhoeven and Vacher (2016). This framework therefore enables treatment of $X_{Fe}$ (or Mg#) as a free parameter. In contrast, Khan et al. (2014) used the best approach available then, in which conductivity is matched with an assemblage of minerals whose individual iron contents were fixed by the original experiments (see citations in **Table A1**). By specifying $X_{Fe}$ at those values, the olivine and garnet conductivities used by Khan and co-workers can be fit reasonably well by the Verhoeven and Vacher model (**Fig. A1**). However, the pyroxenes (measured by Yang et al., 2011; 2012) appear to be abnormally conductive, even allowing for the iron-rich composition of these samples (Mg# ~70). Wang et al. (2016) noted that these measurements were made at the Ni-NO oxygen fugacity (fO$_2$) buffer and suggested a three-fold decrease in conductivity to correct to the more commonly used (and reducing) Mo-MO$_2$ buffer. This brings orthopyroxene (opx) into good agreement with the variable-iron constitutive relation, but the conductivity of the Yang et al. clinopyroxene (cpx) would still be a factor of ~7 higher than that inferred by Wang et al. Compared to the other ferromagnesian minerals, the latter appears to be low, so a multiplier of 7 seems appropriate. Because of this discrepancy, cpx is not used in the main text.

The effect of oxygen fugacity on conductivity was reviewed by Karato and Wang (2013) and was modeled for the Moon by Karato (2013). Small-polaron conductivity increases with fO$_2$ due to increased availability of $Fe^{3+}$. However, the dependence is rather weak, to the ~1/6 power. The Mo-MoO$_2$ redox buffer was used for the experiments leading to the constitutive relations adopted here. This is about +1.5 log units above the Fe-FeO buffer, which in turn is about +0.5 log units above the lunar mantle (Wieczorek et al., 2008; their Fig. 3.17). This difference would lead to a decrease in small-polaron conductivity by about a factor of two. In order to match the

fO$_2$-corrected conductivity, temperature must increase 30–90 K. This small change is neglected in favor of fidelity to the original constitutive relations of Wang et al. (2014) and Verhoeven and Vacher (2016).

Ionic conduction was not separately evaluated by Wang et al. (2014) and Verhoeven and Vacher (2016). At the temperatures and even modest iron content of the Moon, small-polaron conduction dominates over ionic conduction and so the latter can be neglected.

While proton conduction is well-recognized in the literature, the estimated water abundance producing a specified conductivity change is sharply divided. The constitutive relations of Yoshino et al. (2009) and Karato (2013, and references therein) can both be expressed via Eqn. (A1), except that Karato's model includes additional multipliers for oxygen fugacity (fO$_2$/fO$_{2,\text{ref}}$)$^q$. Here again, $q$ varies between dry (small-polaron) and wet (proton) terms, $q_h = 0.17$ and $q_p = -0.1$, respectively. The reference oxygen fugacity fO$_{2,\text{ref}}$ varies as a function of temperature on the Fe-FeO buffer. Rather than attempt to correct over the entire buffer, I sought a normalized oxygen fugacity fO$_2' = $ fO$_2$/fO$_{2,\text{ref}}$ at a reference temperature of 1450 K that would bring Karato's small-polaron conductivities for ol and opx into agreement with Verhoeven and Vacher (2016), and which could then be applied to proton conductivities. These values are fO$_2' = $ 2x10$^7$ for ol and 2x10$^3$ for opx, leading to multipliers (via the $q_h$ power) to $\sigma_0^h$ of 17 and 3.6, respectively. When these fO$_2'$ are taken to the $q_p$ power, the Karato proton $\sigma_0^p$ are multiplied by 0.19 for ol and 0.47 for opx.

**Figure A2** compares the electrical conductivity of olivine computed according to Yoshino et al. (2009) and according to Karato (2013, and references therein), with fO$_2$ normalized for the latter as described above. Karato's results are at a fixed Mg# 83 and so the iron content is fixed as such for the Verhoeven and Vacher approach. At a specified conductivity, then, the Karato formula requires significantly less water than the Yoshino formula.

It should also be noted that the conductivity formula used by Khan et al. (2016) for ilmenite is for a polymorph of (Fe,Mg)SiO$_3$ (Katsura et al., 2007) and not the (Fe,Mg)TiO$_3$ expected to have crystallized late in the solidification of the magma ocean. The conductivity of the latter (Zhang et al., 2006) is two orders of magnitude more conductive than the former. As discussed in the text, however, the net conductivity contribution of low concentrations of ilmenite can be neglected.

# 10. References

Bahr, K., Filloux, J.H., 1989. Local Sq response functions from EMSLAB data. J. Geophys. Res. 94, 14195–14200. https://doi.org/10.1029/JB094iB10p14195

Banks, R.J., 1969. Geomagnetic variations and the electrical conductivity of the upper mantle. Geophys.J.Int. 17, 457–487.

Basaltic Volcanism Study Project (BVSP), 1981. Basaltic Volcanism on the Terrestrial Planets, Pergamon Press, New York, 1286 pp.

Blank, J.L., Sill, W.R., 1969. Response of the Moon to the time-varying interplanetary magnetic field. Journal of Geophysical Research (1896-1977) 74, 736–743. https://doi.org/10.1029/JA074i003p00736.

Carrier, W.D., Olhoeft, G.R., Mendell, W., (eds.G. Heiken, D. Vaniman, and B. French,), 1991. Physical properties of the lunar surface: in Lunar Sourcebook. Cambridge Univ. Press,

Chave, A.D., Jones, A.G., 2012. The Magnetotelluric Method: Theory and Practice. Cambridge University Press.

Constable, S., 2007. 5.07 - Geomagnetism, in: Schubert, G. (Ed.), Treatise on Geophysics. Elsevier, Amsterdam, pp. 237–276.

Constable, S., Parker, R., Constable, C., 1987. Occam's inversion: A practical algorithm for generating smooth models from electromagnetic sounding data. GEOPHYSICS 52, 289–300. https://doi.org/10.1190/1.1442303

Dyal, P., Parkin, C.W., Daily, W.D., 1974. Magnetism and the interior of the Moon. Rev. Geophys. 12, 568–591. https://doi.org/10.1029/RG012i004p00568

Dyal, P., Parkin, C.W., Daily, W.D., 1976. Structure of the lunar interior from magnetic field measurements. Presented at the Lunar and Planetary Science Conference Proceedings, pp. 3077–3095.

Elardo, S.M., Draper, D.S., Shearer, C.K., 2011. Lunar Magma Ocean crystallization revisited: Bulk composition, early cumulate mineralogy, and the source regions of the highlands Mg-suite. Geochimica et Cosmochimica Acta 75, 3024–3045. https://doi.org/10.1016/j.gca.2011.02.033

Elkins-Tanton, L.T., Burgess, S., Yin, Q.-Z., 2011. The lunar magma ocean: Reconciling the solidification process with lunar petrology and geochronology. Earth and Planetary Science Letters 304, 326–336. https://doi.org/10.1016/j.epsl.2011.02.004

Fatemi, S., Fuqua, H.A., Poppe, A.R., Delory, G.T., Halekas, J.S., Farrell, W.M., Holmström, M., 2015. On the confinement of lunar induced magnetic fields. Geophys. Res. Lett. 42, 2015GL065576. https://doi.org/10.1002/2015GL065576

Fuqua Haviland, H., Delory, G.T., de Pater, I., 2019. Finite element analysis for nightside transfer function lunar electromagnetic induction studies. Advances in Space Research 64, 779–800. https://doi.org/10.1016/j.asr.2019.05.006

Gagnepain-Beyneix, J., Lognonné, P., Chenet, H., Lombardi, D., Spohn, T., 2006. A seismic model of the lunar mantle and constraints on temperature and mineralogy. Physics of the Earth and Planetary Interiors 159, 140–166. https://doi.org/10.1016/j.pepi.2006.05.009

Garcia, R.F., Khan, A., Drilleau, M., Margerin, L., Kawamura, T., Sun, D., Wieczorek, M.A., Rivoldini, A., Nunn, C., Weber, R.C., Marusiak, A.G., Lognonné, P., Nakamura, Y., Zhu, P., 2019. Lunar Seismology: An Update on Interior Structure Models. Space Sci Rev 215, 50. https://doi.org/10.1007/s11214-019-0613-y

Grimm, R.E., Delory, G.T., Espley, J.R., McLain, J., Stillman, D.E., 2023. Magnetotelluric sounding of solid-body interiors: Progress and prospects. 54$^{th}$ Lunar Planet. Sci. Conf., #1379.

Grimm, R., Nguyen, T., Persyn, S., Phillips, M., Stillman, D., Taylor, T., Delory, G., Turin, P., Espley, J., Gruesbeck, J., Sheppard, D., 2021. A magnetotelluric instrument for probing the interiors of Europa and other worlds. Advances in Space Research 68, 2022–2037. https://doi.org/10.1016/j.asr.2021.04.011

Grimm, R.E., 2013. Geophysical constraints on the lunar Procellarum KREEP Terrane. J. Geophys. Res. Planets 118, 768–778. https://doi.org/10.1029/2012JE004114

Grimm, R.E., Delory, G.T., 2012. Next-generation electromagnetic sounding of the Moon. Advances in Space Research, Lunar Exploration - I 50, 1687–1701. https://doi.org/10.1016/j.asr.2011.12.014

Hauri, E.H., Saal, A.E., Rutherford, M.J., Van Orman, J.A., 2015. Water in the Moon's interior: Truth and consequences. Earth and Planetary Science Letters 409, 252–264. https://doi.org/10.1016/j.epsl.2014.10.053

Hess, P.C., Parmentier, E.M., 1995. A model for the thermal and chemical evolution of the Moon's interior: implications for the onset of mare volcanism. Earth and Planetary Science Letters 134, 501–514. https://doi.org/10.1016/0012-821X(95)00138-3

Hobbs, B.A., 1977. The electrical conductivity of the Moon: an application of inverse theory. Geophysical Journal International 51, 727–744. https://doi.org/10.1111/j.1365-246X.1977.tb04217.x

Hobbs, B.A., Hood, L.L., Herbert, F., Sonett, C.P., 1983. An upper bound on the radius of a highly electrically conducting lunar core. J. Geophys. Res. 88, B97–B102. https://doi.org/10.1029/JB088iS01p00B97

Hood, L.L., Herbert, F., Sonett, C.P., 1982. The deep lunar electrical conductivity profile: Structural and thermal inferences. J. Geophys. Res. 87, 5311–5326. https://doi.org/10.1029/JB087iB07p05311

Hood, L.L., Mitchell, D.L., Lin, R.P., Acuna, M.H., Binder, A.B., 1999. Initial measurements of the lunar induced magnetic dipole moment using Lunar Prospector Magnetometer data. Geophys. Res. Lett. 26, 2327–2330. https://doi.org/10.1029/1999GL900487

Hood, L.L., Sonett, C.P., 1982. Limits on the lunar temperature profile. Geophys. Res. Lett. 9, 37–40. https://doi.org/10.1029/GL009i001p00037

Johnson, T.E., Morrissey, L.J., Nemchin, A.A., Gardiner, N.J., Snape, J.F., 2021. The phases of the Moon: Modelling crystallisation of the lunar magma ocean through equilibrium thermodynamics. Earth and Planetary Science Letters 556, 116721. https://doi.org/10.1016/j.epsl.2020.116721

Jolliff, B.L., Gillis, J.J., Haskin, L.A., Korotev, R.L., Wieczorek, M.A., 2000. Major lunar crustal terranes: Surface expressions and crust-mantle origins. J. Geophys. Res. 105, 4197–4216. https://doi.org/10.1029/1999JE001103

Karato, S., 2013. Geophysical constraints on the water content of the lunar mantle and its implications for the origin of the Moon. Earth and Planetary Science Letters 384, 144–153. https://doi.org/10.1016/j.epsl.2013.10.001

Karato, S.-I., Wang, D., 2013. Electrical Conductivity of Minerals and Rocks, in: Karato, S.-I. (Ed.), Physics and Chemistry of the Deep Earth. John Wiley & Sons, Ltd, pp. 145–182.

Katsura, T., Yokoshi, S., Kawabe, K., Shatskiy, A., Okube, M., Fukui, H., Ito, E., Nozawa, A., Funakoshi, K., 2007. Pressure dependence of electrical conductivity of (Mg,Fe)SiO3 ilmenite. Phys Chem Minerals 34, 249–255. https://doi.org/10.1007/s00269-007-0143-0

Khan, A., Connolly, J. a. D., Pommier, A., Noir, J., 2014. Geophysical evidence for melt in the deep lunar interior and implications for lunar evolution. Journal of Geophysical Research: Planets 119, 2197–2221. https://doi.org/10.1002/2014JE004661

Khan, A., Connolly, J.A.D., Olsen, N., Mosegaard, K., 2006. Constraining the composition and thermal state of the moon from an inversion of electromagnetic lunar day-side transfer functions. Earth and Planetary Science Letters 248, 579–598. https://doi.org/10.1016/j.epsl.2006.04.008

Kuvshinov, A., Semenov, A., 2012. Global 3-D imaging of mantle electrical conductivity based on inversion of observatory C-responses—I. An approach and its verification. Geophys J Int 189, 1335–1352. https://doi.org/10.1111/j.1365-246X.2011.05349.x

Longhi, J., 2006. Petrogenesis of picritic mare magmas: Constraints on the extent of early lunar differentiation. Geochimica et Cosmochimica Acta, A Special Issue Dedicated to Larry A. Haskin 70, 5919–5934. https://doi.org/10.1016/j.gca.2006.09.023

Mittelholz, A., Grayver, A., Khan, A., Kuvshinov, A., 2021. The Global Conductivity Structure of the Lunar Upper and Midmantle. Journal of Geophysical Research: Planets 126, e2021JE006980. https://doi.org/10.1029/2021JE006980

Moriarty III, D.P., Watkins, R.N., Valencia, S.N., Kendall, J.D., Evans, A.J., Dygert, N., Petro, N.E., 2021. Evidence for a Stratified Upper Mantle Preserved Within the South Pole-Aitken Basin. Journal of Geophysical Research: Planets 126, e2020JE006589. https://doi.org/10.1029/2020JE006589

Nagihara, S., and 5 others, 2022. Heat flow measurements planned for the upcoming robotic missions to Mare Crisium and Schrödinger Basin, Ann. Mtg. Lunar Explor. Analysis Group, #5007.

Neal, C., and 26 others, 2021. The lunar geophysical network missions. Ann. Mtg. Lunar Explor. Analysis Group, #5039.

Olsen, N., 1999. Induction studies with satellite data. *Surveys in Geophysics* 20, 309–40. https://doi.org/10.1023/A:1006611303582.

Ringwood, A.E., 1976. Limits on the bulk composition of the Moon. Icarus 28, 325–349. https://doi.org/10.1016/0019-1035(76)90147-0

Roberts, J.J., Tyburczy, J.A., 1999. Partial-melt electrical conductivity: Influence of melt composition. Journal of Geophysical Research: Solid Earth 104, 7055–7065. https://doi.org/10.1029/1998JB900111

Robinson, K.L., Taylor, G.J., 2014. Heterogeneous distribution of water in the Moon. Nature Geosci 7, 401–408. https://doi.org/10.1038/ngeo2173

Saal, A.E., Hauri, E.H., Cascio, M.L., Van Orman, J.A., Rutherford, M.C., Cooper, R.F., 2008. Volatile content of lunar volcanic glasses and the presence of water in the Moon's interior. Nature 454, 192–195. https://doi.org/10.1038/nature07047

Salmon, J., Canup, R.M., 2012. Lunar Accretion from a Roche-interior fluid disk. ApJ 760, 83. https://doi.org/10.1088/0004-637X/760/1/83

Schubert, G., Cassen, P., Young, R.E., 1979. Subsolidus convective cooling histories of terrestrial planets. Icarus 38, 192–211. https://doi.org/10.1016/0019-1035(79)90178-7

Schubert, G., Schwartz, K., 1972. High-frequency electromagnetic response of the Moon. Journal of Geophysical Research (1896-1977) 77, 76–83. https://doi.org/10.1029/JA077i001p00076

Siegler, M.A., Smrekar, S.E., 2014. Lunar heat flow: Regional prospective of the Apollo landing sites. J. Geophys. Res. Planets 119, 2013JE004453. https://doi.org/10.1002/2013JE004453

Simpson, F., Bahr, K., 2005. Practical Magnetotellurics. Cambridge University Press, Cambridge.

Snyder, G.A., Taylor, L.A., Neal, C.R., 1992. A chemical model for generating the sources of mare basalts: Combined equilibrium and fractional crystallization of the lunar magmasphere. Geochimica et Cosmochimica Acta 56, 3809–3823. https://doi.org/10.1016/0016-7037(92)90172-F

Sonett, C.P., Smith, B.F., Colburn, D.S., Schubert, G., Schwartz, K., 1972. The induced magnetic field of the moon: Conductivity profiles and inferred temperature. Proc. Lunar Planet Sci. Conf. 3rd, pp. 2309–2336.

Toksöz, M.N., Hsui, A.T., Johnston, D.J., 1978. Thermal evolutions of the terrestrial planets. The Moon and the Planets 18, 281–320. https://doi.org/10.1007/BF00896484

Turcotte, D.L., Schubert, G., 1982. Geodynamics - Applications of Continuum Physics to Geological Problems. John Wiley & Sons, New York.

Tyburczy, J.A., 2007. 2.21 - Properties of Rocks and Minerals – The Electrical Conductivity of Rocks, Minerals, and the Earth, in: Schubert, G. (Ed.), Treatise on Geophysics. Elsevier, Amsterdam, pp. 631–642.

Verhoeven, O., Vacher, P., 2016. Laboratory-based electrical conductivity at Martian mantle conditions. Planetary and Space Science 134, 29–35. https://doi.org/10.1016/j.pss.2016.10.005

Vozoff, K., 1991. The magnetotelluric method, in: Electromagnetic Methods in Applied Geophysics (Ed. M. Nabighian). Soc. Explor. Geophys., Tulsa.

Waff, H.S., 1974. Theoretical considerations of electrical conductivity in a partially molten mantle and implications for geothermometry. Journal of Geophysical Research (1896-1977) 79, 4003–4010. https://doi.org/10.1029/JB079i026p04003

Wang, Q., Bagdassarov, N., Xia, Q.-K., Zhu, B., 2014. Water contents and electrical conductivity of peridotite xenoliths from the North China Craton: Implications for water distribution in the upper mantle. Lithos, The lithosphere and beyond: a multidisciplinary spotlight 189, 105–126. https://doi.org/10.1016/j.lithos.2013.08.005

Warren, P.H., Rasmussen, K.L., 1987. Megaregolith insulation, internal temperatures, and bulk uranium content of the moon. J. Geophys. Res. 92, 3453–3465. https://doi.org/10.1029/JB092iB05p03453

Weidelt, P., 1972. The inverse problem of geomagnetic induction. J. Geophys. 38, 257–289.

Whittall, K.P., D.W. Oldenburg, 1992. Inversion of Magnetotelluric Data for a One-Dimensional Conductivity. Society of Exploration Geophysicists, Tulsa.

Wieczorek, M.A., Phillips, R.J., 2000. The "Procellarum KREEP Terrane": Implications for mare volcanism and lunar evolution. J. Geophys. Res. 105, 20417–20430. https://doi.org/10.1029/1999JE001092

Yang, X., Keppler, H., McCammon, C., Ni, H., Xia, Q., Fan, Q., 2011. Effect of water on the electrical conductivity of lower crustal clinopyroxene. Journal of Geophysical Research: Solid Earth 116. https://doi.org/10.1029/2010JB008010

Yang, X., Keppler, H., McCammon, C., Ni, H., 2012. Electrical conductivity of orthopyroxene and plagioclase in the lower crust. Contrib Mineral Petrol 163, 33–48. https://doi.org/10.1007/s00410-011-0657-9

Yoshino, T., 2010. Laboratory Electrical Conductivity Measurement of Mantle Minerals. Surv Geophys 31, 163–206. https://doi.org/10.1007/s10712-009-9084-0

Yoshino, T., Nishi, M., Matsuzaki, T., Yamazaki, D., Katsura, T., 2008. Electrical conductivity of majorite garnet and its implications for electrical structure in the mantle transition zone. Physics of the Earth and Planetary Interiors, Frontiers and Grand Challenges in Mineral Physics of the Deep Mantle 170, 193–200. https://doi.org/10.1016/j.pepi.2008.04.009

Yoshino, T., Matsuzaki, T., Shatskiy, A., Katsura, T., 2009. The effect of water on the electrical conductivity of olivine aggregates and its implications for the electrical structure of the upper mantle. Earth and Planetary Science Letters 288, 291–300. https://doi.org/10.1016/j.epsl.2009.09.032

Zhang, B., Katsura, T., Shatskiy, A., Matsuzaki, T., Wu, X., 2006. Electrical conductivity of $FeTiO_3$ ilmenite at high temperature and high pressure. Phys. Rev. B 73, 134104. https://doi.org/10.1103/PhysRevB.73.134104

Zhang, B.-H., Guo, X., Yoshino, T., Xia, Q.-K., 2021. Electrical conductivity of melts: implications for conductivity anomalies in the Earth's mantle. National Science Review 8, nwab064. https://doi.org/10.1093/nsr/nwab064

Zhang, N., Parmentier, E.M., Liang, Y., 2013. A 3-D numerical study of the thermal evolution of the Moon after cumulate mantle overturn: The importance of rheology and core solidification. Journal of Geophysical Research: Planets 118, 1789–1804. https://doi.org/10.1002/jgre.20121

Zhang, J., Head, J.W., Liu, J., Potter, R.W.K., 2023. Lunar Procellarum KREEP Terrane (PKT) Stratigraphy and Structure with Depth: Evidence for Significantly Decreased Th Concentrations and Thermal Evolution Consequences. Remote Sensing 15, 1861. https://doi.org/10.3390/rs15071861

Zhao, Y., de Vries, J., van den Berg, A.P., Jacobs, M.H.G., van Westrenen, W., 2019. The participation of ilmenite-bearing cumulates in lunar mantle overturn. Earth and Planetary Science Letters 511, 1–11. https://doi.org/10.1016/j.epsl.2019.01.022

## 11. Tables

**Table 1.** Mg# as Functions of Temperature and Mineralogy

| Temperature Model | Mineral | Low-Frequency Fit | All-Frequency Fit | High-Frequency Fit |
|---|---|---|---|---|
| Gagnepain-Beyneix et al., 2006 | ol | 79 ± 3 | 74 ± 9 | 68 ± 8 |
| | opx | 70 ± 3 | 62 ± 14 | 52 ± 10 |
| Grimm, 2013 | ol | 86 ± 2 | 77 ± 14 | 71 ± 14 |
| | opx | 87 ± 7 | 72 ± 23 | 60 ± 20 |
| All | | 81 ± 8 | 71 ± 17 | 63 ± 15 |
| With Conductivity Err. | | 81 ± 10 | 71 ± 19 | 63 ± 17 |

**Table 2.** $H_2O$ (ppmw) at Constant Mg# 81 as Functions of Temperature and Mineralogy

| Temperature Model | Mineral | Low-Frequency Fit | All-Frequency Fit | High-Frequency Fit |
|---|---|---|---|---|
| Gagnepain-Beyneix et al., 2006 | ol[1] | 310 ± 340 | 1100 ± 1100 | 1800 ± 800 |
| | ol[2] | 80 ± 80 | 670 ± 1100 | 1100 ± 1300 |
| | opx[2] | 80 ± 70 | 270 ± 350 | 420 ± 420 |
| Grimm, 2013 | ol[1] | 10 ± 30 | 880 ± 1100 | 1400 ± 1800 |
| | ol[2] | 2 ± 6 | 1300 ± 3300 | 2100 ± 4400 |
| | opx[2] | 3 ± 7 | 420 ± 1000 | 690 ± 1400 |
| All[2] | | 40 ± 60 | 660 ± 1800 | 1100 ± 2400 |
| With Conductivity Err. | | 40 ± 90 | 660 ± 2100 | 1100 ± 2700 |

[1]Yoshino et al., 2009; [2]Karato (2013).

Table A1. Parameters for Electrical Conductivity of Relevant Lunar Mantle Minerals

| Mineral | log $\sigma_0^i$ | $H_i$ | log $\sigma_0^h$ | $H_h$ | $a_h$ | $b_h$ | log $\sigma_0^p$ | $H_p$ | $a_p$ | $b_p$ | Authors |
|---|---|---|---|---|---|---|---|---|---|---|---|
| Olivine | 4.73 | 2.31 | 2.98 | 1.71 | Mg# 92.5 | | 1.90 | 0.92 | 1.0 | 0.16 | Yoshino et al. (2009) |
| Olivine | -- | -- | 2.69 | 2.07 | 2.56 | 0.96 | -- | -- | -- | -- | Wang et al. (2014) |
| Orthopyroxene | -- | -- | 2.39 | 1.09 | Mg# 67 | | 3.83 | 0.84 | -- | 0.90 | Yang et al. (2012) |
| Orthopyroxene | -- | -- | 3.67 | 2.67 | -1.11 | 1.88 | -- | -- | -- | -- | Wang et al. (2014) |
| Clinopyroxene | -- | -- | 2.16 | 1.06 | Mg# 72 | | 3.56 | 0.74 | -- | 1.13 | Yang et al. (2011) |
| Clinopyroxene | -- | -- | 3.20 | 2.74 | -1.11 | 1.88 | -- | -- | -- | -- | Wang et al. (2014) |
| Garnet 900-1300 K<br>1300-1750 K | -- | -- | 1.73<br>3.03 | 1.27<br>1.59 | Mg# 94? | | -- | -- | -- | -- | Yoshino et al. (2008) |
| Garnet | -- | -- | 3.35 | 2.52 | 0 | 1.91 | -- | -- | -- | -- | Verhoeven & Vacher (2016) |
| "Ilmenite" | -- | -- | 1.18 | 0.85 | -- | -- | -- | -- | -- | -- | Katsura et al. (2007) |
| Ilmenite | -- | | 3.35 | 0.21 | -- | -- | -- | -- | -- | -- | Zhang et al. (2006) |

See Eqn. A1 for parameter explanations. First line for each mineral (shaded) was used by Khan et al. (2014). Second line for each mineral (unshaded) tabulated by Verhoeven and Vacher (2016) (except ilmenite) and are used herein for treatment of variable iron fraction.

**12. Figures**

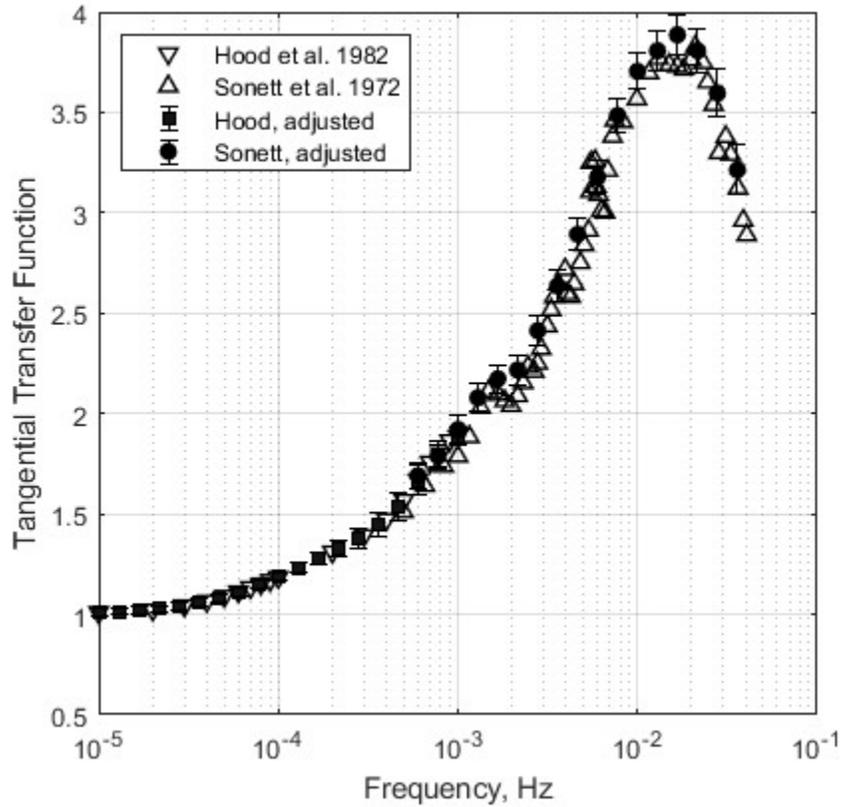

**Figure 1.** Merged datasets for the azimuthal (tangential) component of the Apollo-Explorer magnetic transfer function (TF). The "low-frequency" (LF) data of Hood et al. (1982) is well matched to the high-frequency" (HF) data of *Sonett et al.* (1972) with a shift of +0.1 units to the latter. Final curves are smoothed and interpolated. See text for details. Frequencies up to several mHz can be modeled with dipole response only; higher frequencies require multipoles. Note error bars at lowest frequencies extend to tangential TF<1, which is physically unrealizable.

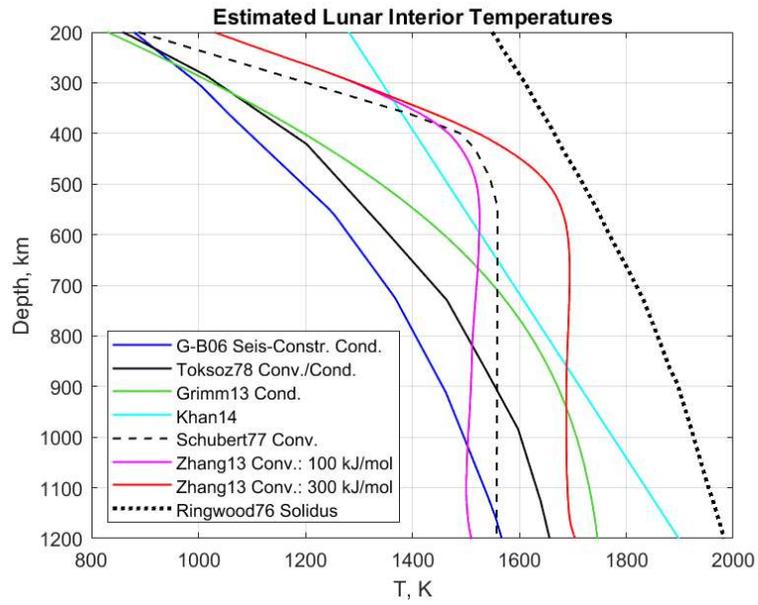

**Fig. 2**. Summary of estimated temperature profiles in the literature. See text for descriptions. Here, electrical conductivity is fit to the conductive temperature profiles of Gagnepain-Beyneix et al. (2006) and Grimm (2013).

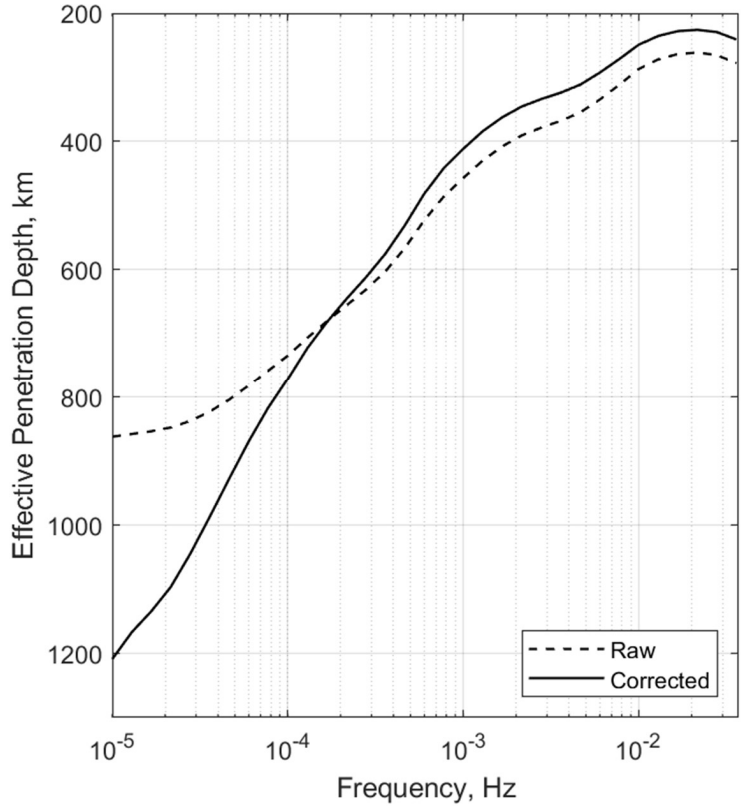

**Figure 3.** Raw and corrected inductive length scales (denoted *C* and *D*, respectively) for Apollo-Explorer merged TF provide one-to-one frequency-depth mapping useful to establish approximate bounds to penetration depth. Raw *C* asymptotes to *a*/2 but remapping by Weidelt transformation and a small empirical adjustment extends over most of the mantle. See text for details. LF data ($10^{-5}$–$10^{-3}$ Hz) span ~400–1200 km depth, whereas HF data ($5\times10^{-4}$–$4\times10^{-2}$ Hz) span ~200-550 km. Dipole approximation completely fails >$2\times10^{-2}$ Hz, yielding erroneous penetration depths.

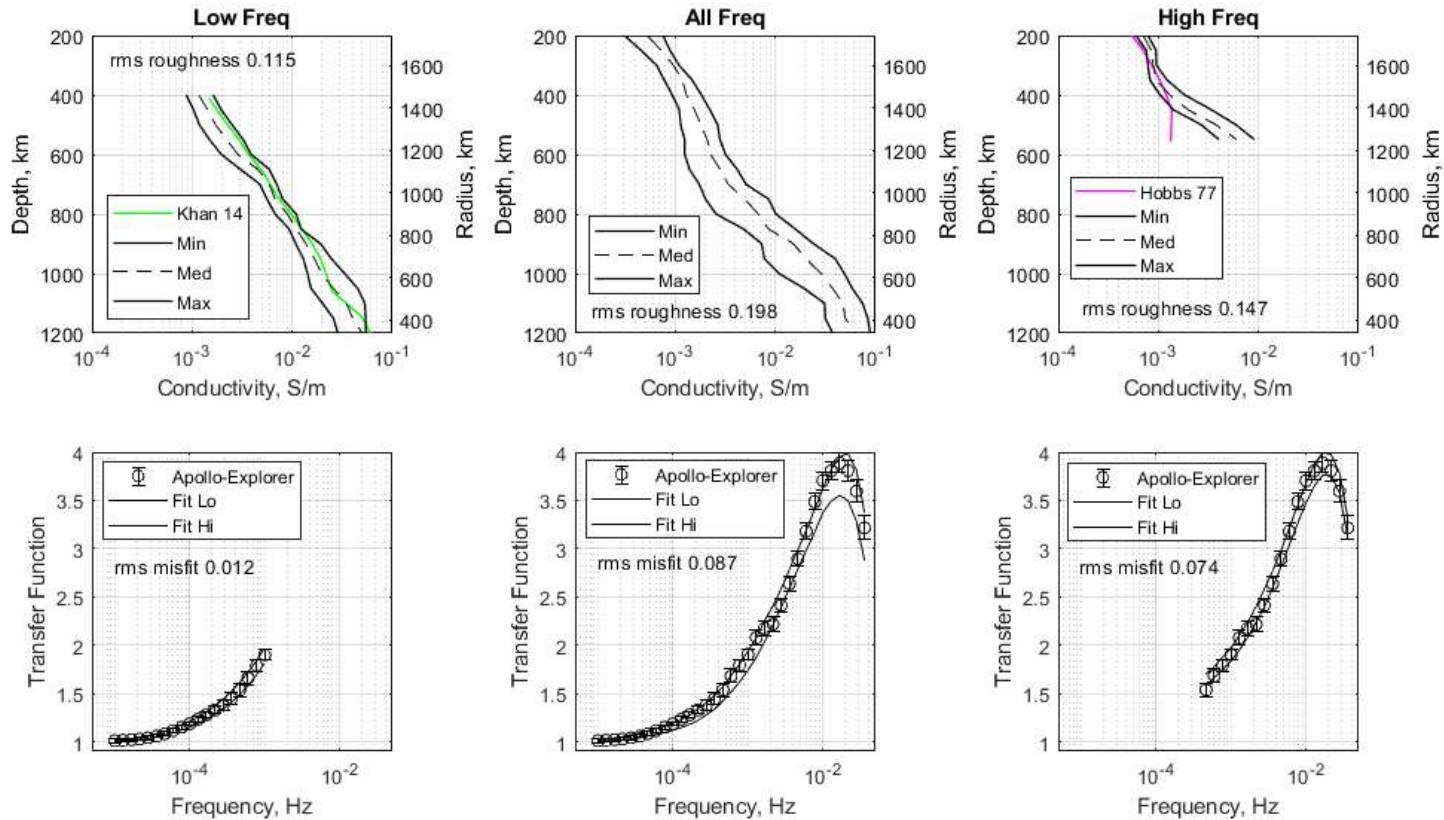

**Figure 4.** Inversion results for low-frequency band (LF, left column), all frequencies (AF, middle column), and high-frequency band (HF, right column). Each column is the result of 45 inversions produced by randomly sampling the observed transfer functions assuming a normal distribution. <u>Top Row</u>: Conductivity-depth solutions, with depth ranges limited according to Fig. 3. Dashed line is median; solid lines bound $68^{th}$ percentile (approximately 1 std. dev.). Note close agreement of LF solution to median of Khan et al. (2014). <u>Bottom Row</u>: Observed transfer functions with standard errors compared to $68^{th}$ percentile of data fits.

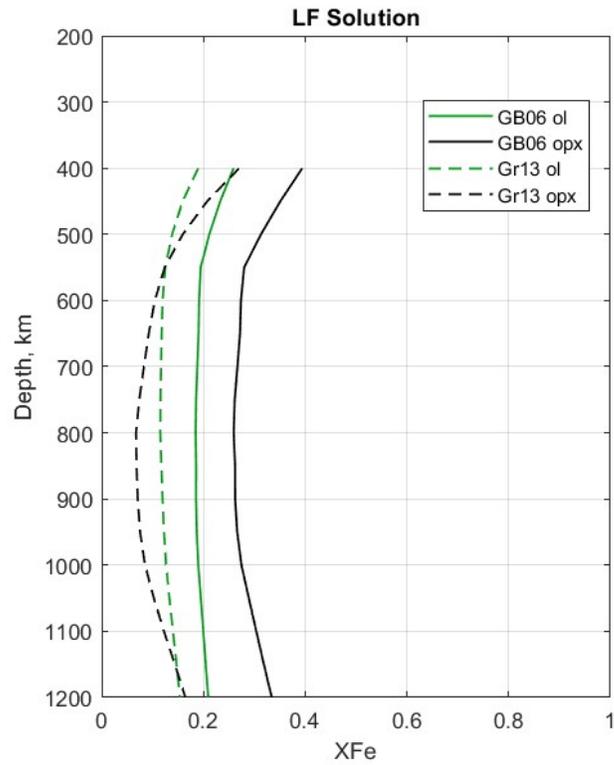

**Figure 5.** Solution for iron fraction $X_{Fe}$ from LF electrical conductivity, using temperature profiles from Gagnepain-Beyneix et al. (2006: GB06) and Grimm (2013; Gr13) and constitutive relations for olivine (ol) and orthopyroxene (opx). Water fraction is zero. See Table 1 for mean Mg# = $100(1-X_{Fe})$ and standard deviation. Vertical variations in iron fraction are small for individual models and influence of temperature is greater than that of composition.

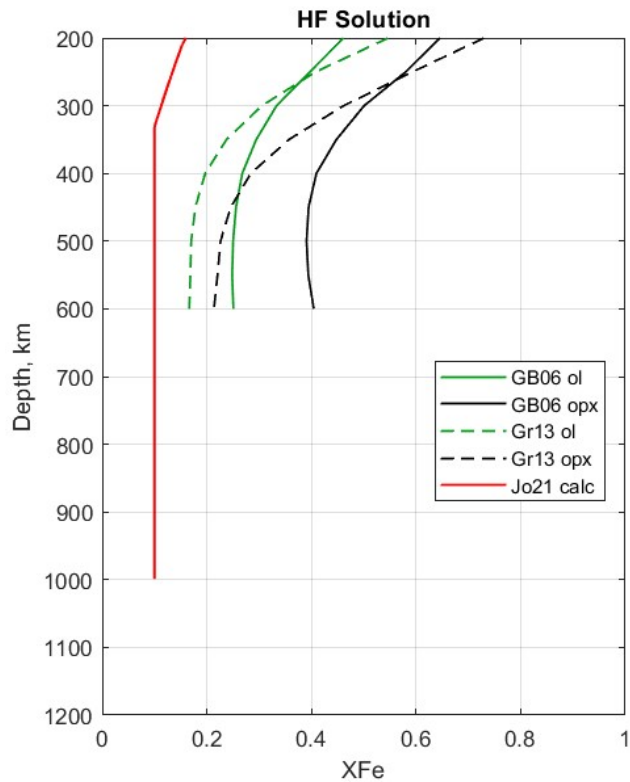

**Figure 6.** As Fig. 5, for fit to HF electrical conductivity. Iron fraction increases strongly with decreasing depth: this may mimic original magma-ocean stratigraphy but is much too large (Johnson et al., 2021: "Jo21" curve). Water can offset the need for high $X_{Fe}$ to match electrical conductivity but would also have to increase with decreasing depth with excessively large mean values >1000 ppmw (Table 2). Trend is likely an artifact, in which HF solution is unable to track strongly decreasing electrical conductivity in the cold upper mantle: this results in erroneously large conductivities and therefore erroneously large contributions of mineral conductivity.

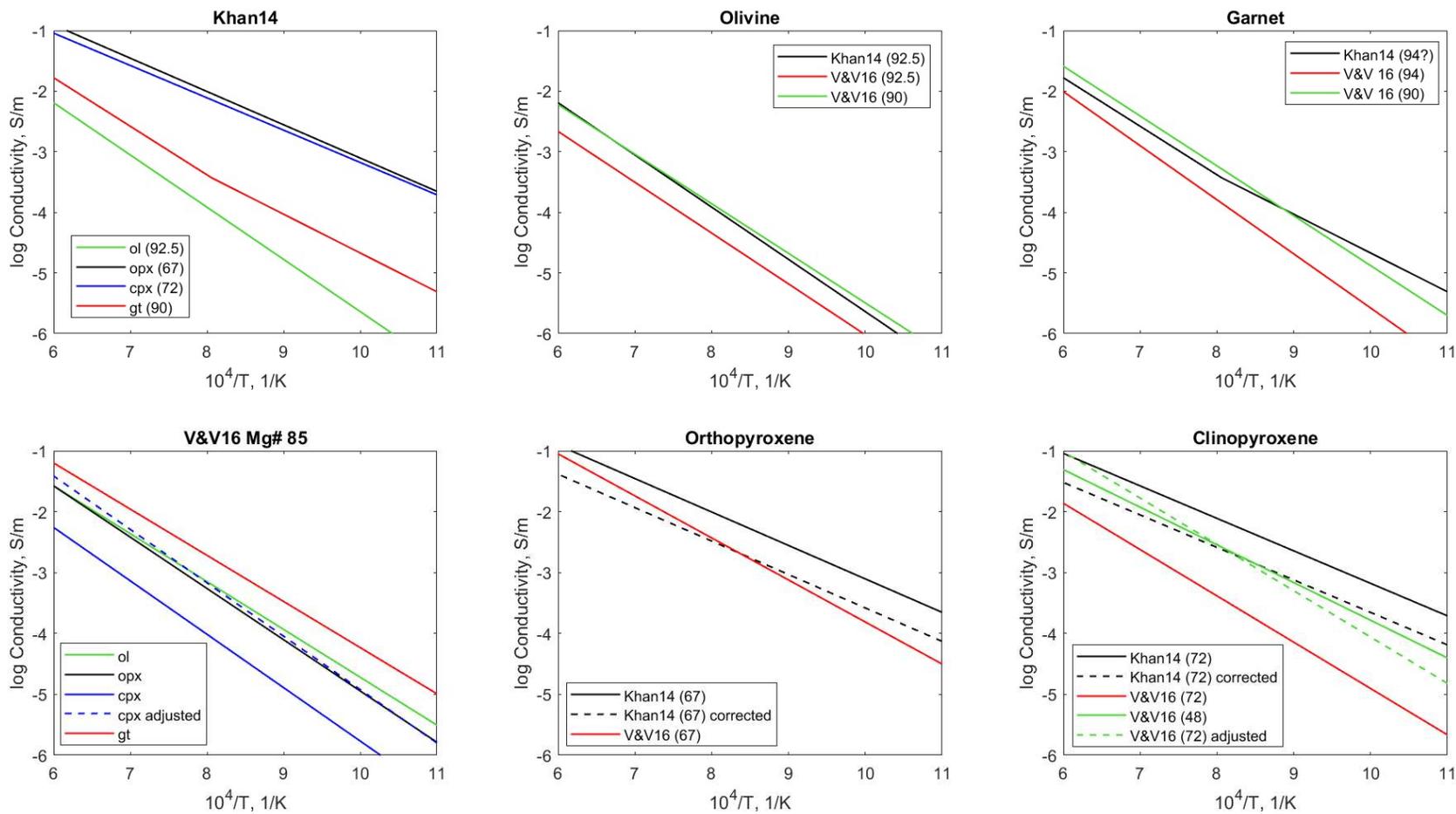

**Figure A1**. Electrical conductivities for lunar-mantle minerals. Khan et al. (2014) used formulae from samples with fixed Mg# (in parentheses next to mineral abbreviations). Pyroxene conductivities are high due to high iron content and high oxygen fugacity. Wang et al. (2014) proposed correction to dashed lines. Formulae in Verhoeven and Vacher (2016) include effects of variable iron content: olivine, garnet, and corrected orthopyroxene agree reasonably well (Mg#s in parentheses) with references cited by Khan and colleagues, but clinopyroxene requires either very low Mg# or an ad hoc conductivity multiplier of 7. "Ilmenite" is $(Mg,Fe)SiO_3$. See Table A1 for original literature citations.

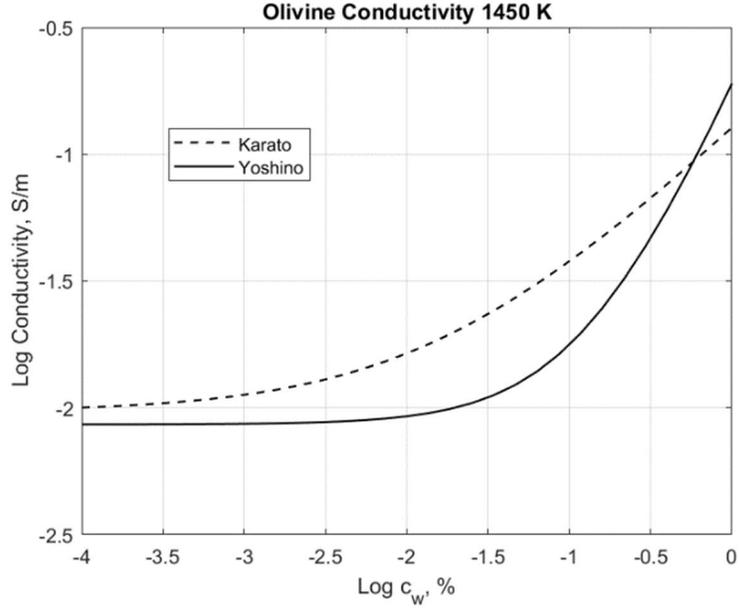

**Fig. A2**. Comparison of effect of water concentration $c_w$ on electrical conductivity, following Yoshino (2009) and Karato (2013, and references therein). The latter's constitutive relation is approximately normalized for $fO_2$ as described in Appendix A. The Yoshino formula is sensitive (>20% increase in conductivity from zero-water value) to $c_w > 2.4 \times 10^{-2}$ % (240 ppmw), whereas the Karato formula is sensitive to $c_w > 1.4 \times 10^{-3}$ % (14 ppmw). Water concentrations (at $c_w < 0.1$%) matching the same conductivity are significantly lower in the Karato formula compared to the Yoshino formula.